\documentstyle[12pt]{article}
 
\makeatletter
 
\if@twoside \oddsidemargin -17pt \evensidemargin 00pt
\else \oddsidemargin 00pt \evensidemargin 00pt
\fi
\let\regeltransport=\baselineskip
\def\section{\@startsection {section}{1}{\z@}%
{ 1\regeltransport plus  .5\regeltransport minus  .25\regeltransport}%
{+0.1pt plus +0.1pt minus +0.1pt}{\reset@font\normalsize\bf}}

\def\subsection{\@startsection{subsection}{2}{\z@}%
{ 1\regeltransport plus  .5\regeltransport minus  .25\regeltransport}%
{+0.1pt plus +0.1pt minus +0.1pt}{\reset@font\normalsize\bf}}

\def\subsubsection{\@startsection{subsubsection}{3}{\z@}%
{ 1\regeltransport plus  .5\regeltransport minus  .25\regeltransport}%
{+0.1pt plus +0.1pt minus +0.1pt}{\reset@font\normalsize\it}}

\def\paragraph{\@startsection{paragraph}{4}{\z@}%
      {3.25ex plus 1ex minus .2ex}{-1em}{\reset@font\normalsize\sl}}
\def\subparagraph{\@startsection{subparagraph}{5}{\parindent}%
      {3.25ex plus 1ex minus .2ex}{-1em}{\reset@font\normalsize\sl}}

\expandafter\ifx\csname mathrm\endcsname\relax\def\mathrm#1{{\rm #1}}\fi
 
\newcount\@tempcntc
\def\@citex[#1]#2{\if@filesw\immediate\write\@auxout{\string\citation{#2}}\fi
  \@tempcnta\z@\@tempcntb\m@ne\def\@citea{}\@cite{\@for\@citeb:=#2\do
    {\@ifundefined
       {b@\@citeb}{\@citeo\@tempcntb\m@ne\@citea
        \def\@citea{,\penalty\@m\ }{\bf ?}\@warning
       {Citation `\@citeb' on page \thepage \space undefined}}%
    {\setbox\z@\hbox{\global\@tempcntc0\csname
b@\@citeb\endcsname\relax}%
     \ifnum\@tempcntc=\z@ \@citeo\@tempcntb\m@ne
       \@citea\def\@citea{,\penalty\@m}
       \hbox{\csname b@\@citeb\endcsname}%
     \else
      \advance\@tempcntb\@ne
      \ifnum\@tempcntb=\@tempcntc
      \else\advance\@tempcntb\m@ne\@citeo
      \@tempcnta\@tempcntc\@tempcntb\@tempcntc\fi\fi}}\@citeo}{#1}}

\def\@citeo{\ifnum\@tempcnta>\@tempcntb\else\@citea
  \def\@citea{,\penalty\@m}%
  \ifnum\@tempcnta=\@tempcntb\the\@tempcnta\else
   {\advance\@tempcnta\@ne\ifnum\@tempcnta=\@tempcntb \else
\def\@citea{--}\fi
    \advance\@tempcnta\m@ne\the\@tempcnta\@citea\the\@tempcntb}\fi\fi}

\makeatother

\def\co{\relax}
\def\co{,}

\def\nlc{\co\nonumber\\}

\def\asymp#1%
{\mathrel{\raisebox{-.4em}{$\widetilde{\scriptstyle #1}$}}}

\def\Nequal#1%
{\mathrel{\raisebox{-.5em}{$\stackrel{=}{\scriptstyle\rm#1}$}}}

\def\beq#1\eeq{\begin{equation}#1\end{equation}}
\def\beqar{\begin{eqnarray}}
\def\eeqar{\end{eqnarray}}
\def\barr#1{\begin{array}{#1}}
\def\earr{\end{array}}
\def\bfi{\begin{figure}}
\def\efi{\end{figure}}
\def\btab{\begin{table}}
\def\etab{\end{table}}
\def\bce{\begin{center}}
\def\ece{\end{center}}
\def\nn{\nonumber}

\def\al{\alpha}
\def\be{\beta}

\def\de{\delta}
\def\De{\Delta}

\def\si{\sigma}

\def\refeq#1{\mbox{(\ref{#1})}}

\def\refta#1{\mbox{Table~\ref{#1}}}

\def\refse#1{\mbox{Section~\ref{#1}}}

\def\citere#1{\mbox{Ref.~\cite{#1}}}
\def\citeres#1{\mbox{Refs.~\cite{#1}}}
 
\newcommand{\TeV}{\unskip\,\mathrm{TeV}}
\newcommand{\GeV}{\unskip\,\mathrm{GeV}}

\newcommand{\pb}{\unskip\,\mathrm{pb}}
\newcommand{\fb}{\unskip\,\mathrm{fb}}

\newcommand{\Oa}{\mathswitch{{\cal{O}}(\alpha)}}

\newcommand{\M}{{\cal{M}}}

\def\mathswitchr#1{\relax\ifmmode{\mathrm{#1}}\else$\mathrm{#1}$\fi}

\newcommand{\PW}{\mathswitchr W}
\newcommand{\PZ}{\mathswitchr Z}

\newcommand{\PH}{\mathswitchr H}
\newcommand{\Pe}{\mathswitchr e}

\newcommand{\Pd}{\mathswitchr d}
\newcommand{\Pu}{\mathswitchr u}
\newcommand{\Ps}{\mathswitchr s}
\newcommand{\Pc}{\mathswitchr c}
\newcommand{\Pb}{\mathswitchr b}
\newcommand{\Pt}{\mathswitchr t}

\newcommand{\Pep}{\mathswitchr {e^+}}
\newcommand{\Pem}{\mathswitchr {e^-}}
\newcommand{\PWp}{\mathswitchr {W^+}}
\newcommand{\PWm}{\mathswitchr {W^-}}

\def\mathswitch#1{\relax\ifmmode#1\else$#1$\fi}

\newcommand{\MW}{\mathswitch {M_\PW}}

\newcommand{\MH}{\mathswitch {M_\PH}}
\newcommand{\Me}{\mathswitch {m_\Pe}}

\newcommand{\Mt}{\mathswitch {m_\Pt}}
 
\newcommand{\sw}{\mathswitch {s_\PW}}

\newcommand{\GF}{\mathswitch {G_\mu}}

\def\ie{i.e.\ }
\def\eg{e.g.\ }

\newcommand{\AAWW}{\gamma\gamma\to\PWp\PWm}

\newcommand{\eeWW}{\Pep\Pem\to\PWp\PWm}

\newcommand{\born}{\mathrm{Born}}

\newcommand{\z}{\setbox0\hbox{+}\hbox to \wd0{\hss0\hss}}

\def\pslash#1{{\setbox0=\hbox{$#1$}
  \rlap{\ifdim\wd0>.7em\kern.22\wd0\else\kern.1\wd0\fi /}#1}}

\def\braket#1#2{\left\langle #1\vphantom{#2}
  \right. \kern-2.5pt\left| #2\vphantom{#1}\right\rangle }

\def\M{{\cal M}}
\def\O{{\cal O}}

\newcommand{\he}{\kappa}

\def\solid{\raise.9mm\hbox{\protect\rule{1.1cm}{.2mm}}}

\newcommand{\app}{\mathrm{app}}

\begin{document}
\thispagestyle{empty}
\def\thefootnote{\fnsymbol{footnote}}
\setcounter{footnote}{1}
\null
\hfill CERN-TH/97-128
\vskip 1cm
\vfil
\begin{center}
{\Large \bf Approximations for W-Pair Production\\ 
at Linear-Collider Energies%
\footnote{Contribution to the Proceedings of the {\it Joint ECFA/DESY
    Study: Physics and Detectors for a Linear Collider,}
Frascati, London, Munich, Hamburg, February 5 to November 22, 1996.}
\par} \vskip 2em
{\large
{\sc A.\ Denner$^1$ and S.~Dittmaier$^2$} \\[10ex]
\parbox{12cm}{\normalsize
{\it$^1$ PSI, W\"urenlingen und Villigen, Switzerland} \\[1ex]
{\it$^2$ CERN, Theory Division, Geneva, Switzerland} 
}
\par} \vskip 1em
\end{center} \par
\vskip 2cm
\vfil
{\bf Abstract} \par
We determine the accuracy of various approximations to the $\Oa$
corrections for on-shell \PW-pair production. While an approximation
based on the universal corrections arising from initial-state
radiation, 
from the running of $\alpha$, and 
from corrections proportional to $\Mt^2$
fails in the Linear-Collider 
energy range, a high-energy approximation 
improved by the exact universal corrections
is sufficiently good above about $500\GeV$. These results indicate that in
Monte Carlo event generators 
for off-shell \PW-pair production the
incorporation of the universal corrections is not sufficient and more
corrections should be included. 
\par
\vskip 1.5cm
\noindent CERN-TH/97-128 
\par
\vskip .15mm
\noindent June 1997 \par
\null
\setcounter{page}{0}
\clearpage
 
\def\thefootnote{\fnsymbol{footnote}}
\setcounter{footnote}{1}
\null
\vskip 2.0em  minus 0.2em
\begin{center}
{\Large \bf Approximations for W-Pair Production\\
at Linear-Collider Energies%
\par} \vskip 2.0em  minus 0.5em
{\large
{\sc A.\ Denner$^2$ and S.~Dittmaier$^2$} \\[3.5ex]
\parbox{11cm}{\normalsize
{\it$^1$ PSI, W\"urenlingen und Villigen, Switzerland} \\[.8ex]
{\it$^2$ CERN, Theory Division, Geneva, Switzerland} 
}
\par}
\end{center} \par
\vskip 3.3em plus 1em minus 1em
 
\renewcommand{\thefootnote}{\arabic{footnote}}
\setcounter{footnote}{0}
\section{Introduction}
\label{SEintro}
 
One of the most important processes for testing 
the Minimal Standard Model (MSM) in the future is the production of \PW\ 
pairs \cite{Be94,WWgeneral,EE500gen}.  It allows a direct study of the
non-Abelian triple gauge couplings in the clean environment of
$\Pep\Pem$ collisions.  
The sensitivity of this process to anomalous gauge-boson couplings
grows strongly with energy owing to the fact that these couplings in
general spoil the unitarity cancellations that are present in the
MSM for longitudinal gauge bosons. Consequently, more stringent limits
on non-standard couplings can be obtained at higher energies.
 
In order to test for anomalous couplings,
the cross-section for \PW-pair production has to be known
with an accuracy of $1\%$ or better. At this level, the inclusion
of radiative corrections is mandatory. However, so far the complete
electroweak $\Oa$ corrections to off-shell \PW-pair production are not
available and the present 
Monte Carlo event generators 
include only the known leading universal corrections.

In order to assess the theoretical uncertainty inherent in these 
event generators,
on-shell \PW-pair production can be used as guideline. The
corresponding cross-section in the MSM is known, including the
complete set of electroweak $\Oa$ corrections \cite{eeww}.
The size of the non-leading $\Oa$
corrections in the on-shell case should provide a reasonable estimate for the
corresponding left-out non-leading corrections in the off-shell case.

In this short article we discuss the quality of an improved Born
approximation (IBA) for on-shell \PW-pair production that is based on
the known leading universal corrections. We concentrate on the 
Linear-Collider (LC)
energy range and emphasize the differences to the LEP2 case studied in
\citere{lep295}.  In addition, we compare the full one-loop results
with a form-factor approximation (FFA), which corresponds to the best
possible improved Born approximation, and with a 
consistent high-energy approximation (HEA).
We also include some remarks on the radiative corrections to \PW-pair
productions in photon--photon collisions.

\section{Form-factor approximation}
\label{SEffa}

Whereas in the lowest-order matrix element for $\eeWW$ only three
different tensor structures occur, at $\Oa$ twelve independent tensor
structures are required, each of which is associated with an
independent invariant function. The dominant radiative corrections, 
such as those that are related to
UV, IR or mass singularities, in general have factorization properties
and are at \Oa\ restricted to those invariant functions that appear
already at lowest order. Therefore, the contributions of the other
invariant functions should be relatively small.  Indeed, a numerical
analysis reveals that in a suitably chosen representation for the
basic set of independent matrix elements only the three Born-like
invariant functions plus one extra right-handed piece are relevant for
a sufficiently good approximation \cite{Di92,Fl92}.
This suggests an approximation for the matrix element of the form
\beq \label{Mapp} 
\M^{\he}_{\app} = \M^{\he}_{I} F_I^{\he} + \M^{\he}_{Q} F_Q^{\he}, 
\eeq 
where---following the notation of \citere{Be94}---$\M_I$ and $\M_Q$ 
denote the tensor structures associated with
the charged-current coupling and the electromagnetic coupling,
respectively, in the lowest-order matrix element, and $\he$ is 
the electron helicity. The term involving $F_I^+$ is only needed for
right-handed electrons in the 
LEP2 energy region.  Neglecting the
other invariant functions in the basis 
of \citere{Di92} defines
the FFA. The form factors $F_{I}^{\he}$ and $F_Q^{\he}$ are independent of
the \PW~polarizations and include all corrections that are related to
the Born structure. However, they are both energy- and
angular-dependent and their evaluation is as complicated as the exact
one-loop calculation. 
 
\section{Improved Born approximation}
\label{SEiba}

In order to construct an IBA one has to specify
simple expressions that reproduce the invariant functions
$F_{I,Q}^\he$ with sufficient accuracy. 
In the LEP2 energy region the following expressions provide 
a reasonable ansatz \cite{Di92}
\beqar \label{FIapp}
\nn F_{I,\mathrm{IBA}}^\he &=& \left[
2\sqrt{2}\GF\MW^2+ \frac{4\pi\al}{2\sw^2}\frac{\pi\al}{4\beta}
(1-\beta^2)^2\right]\delta_{\he-} \nlc[1ex]
F_{Q,\mathrm{IBA}}^\he &=& \left[
4\pi\al(s)+ 4\pi\al\frac{\pi\al}{4\beta}
(1-\beta^2)^2 \right].
\eeqar
As usual, $s$, $t$, and $u$ denote the 
Mandelstam variables, and $\be=\sqrt{1-4\MW^2/s}$ is the W-boson
velocity in the 
centre-of-mass system.
 
The terms containing $\GF$ and $\al(s)$ incorporate all 
leading universal corrections associated with the running of $\al$
and the corrections $\propto \al\Mt^2/\MW^2$ associated with the 
$\rho$ parameter.
As these are linked to the renormalization of the electric charge at zero
momentum transfer and of the weak mixing angle,
they only contribute to the structures already present at lowest order.
The $1/\beta$ term describes the effect of the Coulomb singularity,
which is only relevant close to threshold and can be omitted at high
energies. The factor $(1-\be^2)^2$ is introduced by hand to restrict the
$1/\be$ contribution to the threshold region. 
The ansatz \refeq{FIapp} includes all leading logarithms originating
from light fermions and all corrections 
$\propto \al\Mt^2/\MW^2$.
As has been studied in \citere{Di92}, contributions 
$\propto\alpha\log(\Mt/\MW)$ and $\propto\alpha\log(\MH/\MW)$ 
that occur when the masses are heavy and 
are not covered by \refeq{FIapp} have a negligible numerical impact.
 
In addition to the contributions described so far, one has to include
the leading logarithmic QED corrections. These can be calculated using
the structure-function method \cite{Ku85}.
They comprise all contributions $\propto\al\log(\Me/\MW)$.
Since we are interested in approximations for the virtual corrections,
we restrict ourselves to the inclusion of real photonic corrections in
the soft-photon approximation. In this approximation the leading
logarithmic QED corrections to the cross-section are given by the factor
\beqar\label{IBA}
\de_\gamma &=&
2\frac{\al}{\pi}\biggl[\,
\ln\frac{s}{\Me^2}\ln\frac{2\De E}{\sqrt{s}}
+ \frac{3}{4}\ln\frac{Q^2}{\Me^2}
\nn\\
&& \phantom{2\frac{\al}{\pi}\biggl[\,}
- \ln\frac{2\De E}{\sqrt{s}}\biggl(2-2\ln\frac{\MW^2-t}{\MW^2-u}
- \frac{s-2\MW^2}{s\be}\ln\frac{1+\be}{1-\be}\biggr)\,\biggr],
\eeqar
where $\De E$ is the soft-photon cutoff energy, 
and $Q^2$ is a scale that is not determined in 
the leading-log approximation and set to $Q^2=s$ for the numerics.

We note that all present event generators for off-shell \PW-pair production
include only those radiative corrections that are included in our IBA.

\section{High-energy approximation}
\label{SEhea}

At high energies the full $\Oa$ corrections can be approximated 
by exploiting the fact that $s, |t|, |u| \gg M_{\PW,\PZ}^2\gg
m_{\Pe,\mu,\tau,\Pd,\Pu,\Ps,\Pc,\Pb}$. Such an approximation has been
constructed for the process $\Pep\Pem\to\PWp\PWm$ by a systematic
expansion of the exact one-loop corrections for arbitrary \Mt\ and
\MH\ \cite{wwhe,wwhel}.  For intermediate energies ($500\GeV$ to
$2\TeV$) the high-energy approximation is improved \cite{wwhe} by
exactly taking into account the leading universal corrections of the
IBA, defined in \refse{SEiba}, apart from the Coulomb singularity. As
the HEA has already been discussed in \citere{eeWWhep}, we summarize
here only its most important features.
 
The dominant corrections at high energies result from 
terms proportional to the high-energy leading logarithms
$\alpha\log(q_1^2/\MW^2)\log(q_2^2/\MW^2)$ with $q_i^2=s, |t|, |u|$.
These corrections are non-universal and arise from vertex and box 
diagrams.
For longitudinal gauge bosons, additional sizeable corrections appear,
depending on the values of the top-quark and Higgs-boson masses.  In
the limit 
$s \gg \Mt^2,\MH^2 \gg \MW^2$, terms containing
$\alpha\Mt^2/\MW^2\log(\Mt/\MW)$ or $\alpha\MH^2/\MW^2$ arise as a
consequence of incomplete screening.  In the limit 
$\Mt^2,\MH^2 \gg s \gg \MW^2$, corrections of the form 
$\al s/\MW^2\log(\Mt/\MW)$ and $\al s/\MW^2\log(\MH/\MW)$ result
from delayed unitarity effects.
 
\section{Numerical results}
\label{SEnures}
 
For the numerical evaluation we used the parameters given in
\citere{lep295}. The corresponding values for the \PW-boson,
Higgs-boson, and
top-quark masses are $\MW=80.26\GeV$, $\MH=300\GeV$, and $\Mt=165.3\GeV$.

The unpolarized lowest-order cross-sections integrated over the
angular region $10^\circ\le\theta \le170^\circ $ and the differential
cross section for three angles are listed in \refta{quality} at various
energies.  The quality $\Delta_{\mathrm{app}}$ is defined as 
\beq
\Delta_{\mathrm{app}} = 
\left|\frac{\de\sigma_{\mathrm{app}} - \de\si_{\mathrm{exact}}}
{\si^{\born}_{\mathrm{exact}}}\right|  ,
\eeq
where $\de\si_{\mathrm{exact}}$ is the full virtual and soft-photonic
$\Oa$ correction, $\de\si_{\mathrm{app}}$ the corresponding
approximation and $\si^{\born}_{\mathrm{exact}}$ the exact
lowest-order cross-section.
When calculating the difference $\de\sigma_{\mathrm{app}} -
\de\si_{\mathrm{exact}}$, 
care has to be taken that 
the leading higher-order corrections are treated in the same way
in the two expressions.
\btab
\bce
{\normalsize
\begin{tabular}{|c|c||r|r|r|r|}
\hline
$\sqrt{s}/\GeV$ & $\theta$ & $\si_{\mathrm{B}}/\fb$ & 
$\De_{\mathrm{IBA}}/\%$ & $\De_{\mathrm{FFA}}/\%$ & $\De_{\mathrm{HEA}}/\%$ \\
\hline\hline
$161$ & $(10^\circ,170^\circ)$ & 
$3753.2$ & $1.5\phantom{0}$ & $0.00$ & $37\phantom{.00}$ \\
\cline{2-6}
&  $10^\circ$ & 
$367.0$ & $1.6\phantom{0}$ & $0.00$ & $36\phantom{.00}$ \\
\cline{2-6}
&  $90^\circ$ & 
$300.7$ & $1.4\phantom{0}$ & $0.00$ & $37\phantom{.00}$ \\
\cline{2-6}
& $170^\circ$ & 
$250.0$ & $1.3\phantom{0}$ & $0.00$ & $37\phantom{.00}$ \\
\hline
$175$ & $(10^\circ,170^\circ)$ & 
$15591$ & $1.3\phantom{0}$ & $0.03$ & $12\phantom{.00}$ \\
\cline{2-6}
&  $10^\circ$ & 
$3380$ &  $1.7\phantom{0}$ & $0.00$ & $10\phantom{.00}$ \\
\cline{2-6}
&  $90^\circ$ & 
$1001$ &  $1.0\phantom{0}$ & $0.05$ & $12\phantom{.00}$ \\
\cline{2-6}
& $170^\circ$ & 
$439$ & $0.59$ & $0.00$ & $12\phantom{.00}$ \\
\hline
$200$ & $(10^\circ,170^\circ)$ & 
$17107$ & $1.5\phantom{0}$ & $0.01$ & $3.7\phantom{0}$ \\
\cline{2-6}
&  $10^\circ$ & 
$6463$ & $1.8\phantom{0}$ & $0.00$ & $2.3\phantom{0}$ \\
\cline{2-6}
&  $90^\circ$ & 
$812$ & $1.4\phantom{0}$ & $0.02$ & $4.7\phantom{0}$ \\
\cline{2-6}
& $170^\circ$ & 
$255$ & $1.3\phantom{0}$ & $0.00$ & $3.8\phantom{0}$ \\
\hline
$500$ & $(10^\circ,170^\circ)$ & 
$4413.1$ & $4.7\phantom{0}$ & $-0.06$ & $-0.85$ \\
\cline{2-6}
&  $10^\circ$ & 
$11604.4$ & $1.9\phantom{0}$ & $0.00$ & $-0.67$ \\
\cline{2-6}
&  $90^\circ$ & 
$75.4$ & $10\phantom{.00}$ & $-0.29$ & $-0.05$ \\
\cline{2-6}
& $170^\circ$ & 
$6.5$ & $14\phantom{.00}$ & $-0.19$ &  $3.5\phantom{0}$ \\
\hline
$1000$ & $(10^\circ,170^\circ)$ & 
$1084.3$ & $11\phantom{.00}$ & $0.06$ & $0.21$ \\
\cline{2-6}
&  $10^\circ$ & 
$3292.3$ & $3.9\phantom{0}$ & $0.00$ & $1.1\phantom{0}$ \\
\cline{2-6}
&  $90^\circ$ & 
$16.7$ & $23\phantom{.00}$ & $0.08$ & $0.54$ \\
\cline{2-6}
& $170^\circ$ & 
$0.6$ & $28\phantom{.00}$ & $-0.77$ & $6.4\phantom{0}$ \\
\hline
$2000$ & $(10^\circ,170^\circ)$ & 
$267.57$ & $22\phantom{.00}$ & $0.12$ & $0.17$ \\
\cline{2-6}
&  $10^\circ$ & 
$823.35$ & $9.7\phantom{0}$ & $0.02$ & $0.64$ \\
\cline{2-6}
&  $90^\circ$ & 
$4.03$ & $39\phantom{.00}$ & $-0.16$ & $0.34$ \\
\cline{2-6}
& $170^\circ$ & 
$0.09$ & $46\phantom{.00}$ & $-2.3\phantom{0}$ & $5.4\phantom{0}$ \\
\hline
\end{tabular}
}
\ece
\caption{Quality of the approximations for various energies}
\label{quality}
\etab
 
The FFA is excellent; the corresponding error is well below the
per-cent level, whenever the cross-section is sizeable. At high
energies and large scattering angles, where the cross section is
extremely small, larger differences appear. This signals a strong
dominance of the Born structure and demonstrates that improved Born
approximations are possible.  Above LEP2 energies the IBA is
reasonable only for extremely small scattering angles, and deviates
for most angles by several per cent
easily.  This shows that
approximations based on the leading universal corrections are certainly not
sufficient for 
a comparison with experiment at high energies.  
The HEA on the other hand is good
at energies above about $500\GeV$ for sizeable cross-sections,
\ie its deviation from the full $\Oa$ corrections is below 1\%.

\section{Remarks on \boldmath{$\AAWW$}}

Another important process for the study of anomalous gauge-boson couplings
is \PW-pair production in photon--photon collisions. The corresponding
cross-section approaches $80\pb$ at high energies and is thus much
larger than the one for  \PW-pair production in $\Pep\Pem$ collisions. 
The corrections to the corresponding on-shell process have been studied in
\citeres{aaww} and briefly summarized in \citere{aawwproc}.
Almost all leading universal corrections are absent.
There are no effects from the running of 
$\alpha$, no corrections proportional to $\Mt^2$, and no 
enhanced logarithms arising from soft or collinear photons. 
Not even corrections $\propto\alpha\log(\MH/\MW)$ 
arise in the limit of a large Higgs-boson mass. 
The Coulomb singularity is unimportant at LC energies. Thus, 
an IBA that is equivalent to the one discussed for $\eeWW$ at
high energies becomes trivial, \ie equal to the Born approximation. 
All dominating corrections are of non-universal origin and 
their size, which reaches several per cent at $500\GeV$ and increases
with energy, indicates that they cannot be neglected. 
Unlike for $\eeWW$, no form-factor and high-energy approximations
have been worked out for $\AAWW$ so far.

\section{Conclusions}

At high energies the $\O(\alpha)$ corrections to on-shell W-pair
production 
in $\Pep\Pem$ collisions are large 
[$\O(10$--$40\%)$] and cannot be neglected.
They are not dominated by universal corrections from initial-state
radiation, from the running of 
$\alpha$, and from corrections proportional to
$\Mt^2$, but by contributions like $\alpha\log^2(s/\MW^2)$, which 
originate from the bosonic (vertex and box) corrections. 
For on-shell W-pair production 
they are well approximated by a consistent high-energy expansion that is 
improved by the exact leading universal corrections in
order to yield a sufficient accuracy at intermediate energies.

Present event generators for off-shell W-pair
production include only the universal corrections. The theoretical
uncertainty from neglecting non-universal corrections strongly
increases with energy. While this uncertainty is only 
$1$--$2\%$ at LEP2 energies it amounts to several 
tens of per cent in the TeV range.
Thus, the non-universal corrections are required for adequate
theoretical predictions.
This applies in the same way to $\AAWW$ where essentially
all non-universal corrections are absent.

A first step towards the inclusion of non-universal corrections into
Monte Carlo event generators for $\Pep\Pem\to 4$ fermions is provided
by the fermion-loop scheme \cite{bhf2}, where gauge-boson widths are
consistently introduced by the resummation of the fermion-loop
corrections.  The incorporation of the bosonic corrections, which
dominate the non-universal corrections at high energies, is more
complicated owing to their complexity and problems with gauge
invariance. The most promising approach consists in an expansion
according to the degree of resonance. In this approach the corrections
are to a large extent given by the known corrections for on-shell
W~bosons or, at LC energies, by the corresponding high-energy approximation.
For a genuine on-shell approach to W-pair production and
the subsequent W~decay, the ${\cal O}(\alpha)$ corrections have
already been included in an event generator (see \eg \citere{Ja97}). The
combination of off-shell effects and ${\cal O}(\alpha)$ corrections is
still under investigation.

\end{document}